\documentclass[preprint,aps,prb,floatfix,a4paper,showpacs]{revtex4}
\usepackage{amsfonts}
\usepackage{amssymb}
\usepackage{graphicx}
\usepackage{floatflt}
\usepackage{graphics}

\makeatletter
\newcommand{\row}[1]{\mathord{\buildrel{\lower3pt\hbox{$\scriptscriptstyle\rightarrow$}}\over #1}}

\newcommand{\dyadic}[1]{\mathord{\dyadic@rrow{#1}}}
\newcommand{\dyadic@rrow}[1]{
\begin{picture}(12,12)(-1,0)
\put(-3,12){\makebox(0,0)[t]{$\scriptscriptstyle\downarrow$}}
\put(-3,13){\makebox(0,0)[l]{$\scriptscriptstyle\longrightarrow$}}
\put(5,0){\makebox(0,0)[b]{$#1$}}
\end{picture}
}

\input{tcilatex}
\begin{document}

\title{Information under Lorentz transformation}
\author{ N. Metwally$^{1,2}$, H. Eleuch$^{3}$ and M. Abdel-Aty$^{2,4,5}$ }
\affiliation{$^{1}$Department of Mathematics, Faculty of Science,
Aswan University,
Aswan, Egypt \\
$^{2}$Mathematics Department, College of Science, Bahrain University, Bahrain%
\\
$^{3}$DIRO, Universit\'{e} de Montr\'{e}al, H3T 1J4, Montr\'{e}al, Canada \\
$^{4}$Scientific Publishing Center, Bahrain University, Bahrain \\
$^{5}$Department of Mathematics, Faculty of Science, Sohag
University, Sohag, Egypt }

\date{\today }

\begin{abstract}
A general form of a two-qubit system is obtained under the effect of Lorentz
transformation. We investigate extensively some important classes in the
context of quantum information. It is shown Lorentz transformation causes a
decay of entanglement and consequently information loses. On the other hand,
it generates entangled states between systems prepared initially in a
separable states. The partial entangled states are more robust under Lorentz
transformation than maximally entangled states. Therefore the rate of
information lose is larger for maximum entangled states compared with that
for partially entangled states.

\end{abstract}

\maketitle

\section{Introduction}

It is now well known that the increase of the classical computers
speed has physical limitations \cite{chen}. These limitations are
fundamentally encrypted in the quantum mechanical effects. Since
few decades several groups are developing a new concept of
information processing, quantum information processing, in order
to overcome the classical information processing limitations
\cite{chen,Nielsen}. One of the most powerful tools for the
quantum computing is the entanglement, a pure quantum effect
allowing to speed up the quantum algorithms and to exchange the
information in non classical way. In the last two decades a large
number of works has been done to study the entanglement in
physical systems \cite{Xiao,Eskandari,Soares,el1,aty1,kim02}.
However almost all these contributions were limited to the
non-relativistic effects. Since the development of the special
relativity \cite{Einstein} the way of looking to
the dynamical systems with high speed is drastically changed. Dirac \cite%
{Dirac,Dirac2} introduced the relativistic effect in quantum mechanics just
few years of the establishing its concepts and foundations.

Recently there are some achievements on relativistic quantum
information. For example, Saldanha and Vedral \cite{Saldanaha2012}
 show that a massive 2- qubit particles prepared in a maximally
 entangled state is not capable of maximally violating the
 Clouser-Horne Shimony-Holt inequality.
A scheme for storing quantum information in the field modes of
moving cavities in non-inertial frames was reported \cite{Downes}.
Choi  investigated the relativistic effects on the spin of
entanglement of two massive Dirac particles \cite{Choi2011}. The
effect of the special relativity on the entanglement between spins
and momenta of two-qubit system is investigated by Cafaro et. al.
\cite{Cafaro2012}. Some properties of a system of two-spin -1
particles under Lornetz transformation have investigated  by Ruiz
and E. N.-Achar \cite{Esteban2012}

In this work we study, in the relativistic context, the dynamical behavior
of the entanglement for some particular and important classes of initial
states of two-qubit system, namely Werner, $x-$ and generic pure states. We
analyze the dynamical evolution of the entropy and the information loses for
this system. We show that non-intuitive results are emerging from the
relativistic transformation.

The paper is organized as follows. In Sec. 2, a short review of the effect
of Lorentz transformation on a two qubits state is presented. In Sec. 3, we
describe the couplings between the two qubits, and then we obtain the Bloch
vectors under the effect of the lorentz transformation. In Sec. 4, we
discuss how to characterize the quantum entanglement by using the
concurrence, in contrast to the Werner state and Bell states. In Sec. 5, the
basic principle of information loss is discussed. In particular, we discuss
the effect of Lornetz transformation on the local and non local information
via calculating the entropy of both subsystems and the total state. Finally,
we summary the main results of the paper in Sec. 6.

\section{Model Description and Lorentz Transformation}

In this section we review the effect of Lorentz transformation on a two
qubits state given in the rest frame as:
\begin{equation}
\bigl|\Psi \bigr\rangle=\bigl|\psi _{p_{a}}\bigr\rangle\bigl|\psi _{p_{b}}%
\bigr\rangle\bigl|\psi _{s}\bigr\rangle,  \label{ms}
\end{equation}%
where$\bigl|\psi _{p_{a}}\bigr\rangle$ and $\bigl|\psi _{p_{b}}\bigr\rangle$
are the momentum states for the first and the second qubit respectively,
while $\bigl|\psi _{s}\bigr\rangle$ represents the spin state vector for
both particles \cite{Hui}. A Lorentz transformation $\Lambda $ changes $%
\bigl|\psi _{p_{i}}\bigr\rangle$ to $\bigl|\Lambda \psi _{p_{i}}\bigr\rangle$%
, where $i=a,b$. This transformation represents a unitary operator on the
space of momenta space\cite{Jordan}. Therefore the Lorentz transformation
change the state(\ref{ms}) as:
\begin{equation}
\bigl|\Psi ^{\prime }\bigr\rangle=\bigl|\Lambda \psi _{p_{a}}\bigr\rangle%
\bigl|\Lambda \psi _{p_{b}}\bigr\rangle\mathcal{U}_{a}(\psi _{pa})\mathcal{U}%
_{b}(\psi _{pb})\bigl|\psi _{s}\bigr\rangle,
\end{equation}%
where,$\mathcal{U}_{a}$ and $\mathcal{U}_{b}$ are operators on the spin
states for both particles. In the computational basis $\bigl|00\bigr\rangle,%
\bigl|01\bigr\rangle,\bigl|10\bigr\rangle$ and $\bigl|11\bigr\rangle$, the
operator $\mathcal{U}=\mathcal{U}_{a}\otimes \mathcal{U}_{b}$ can be written
as
\begin{equation}
\mathcal{U}=\bigl|00\bigr\rangle\bigl\langle00\bigr|+\bigl|11\bigr\rangle%
\bigl\langle11\bigr|+e^{-i\phi }\bigl|01\bigr\rangle\bigl\langle10\bigr|%
+e^{i\phi }\bigl|10\bigr\rangle\bigl\langle10\bigr|.
\end{equation}

Assume that a source supplies two users Alice and Bob with a two qubits of
massive Dirac particles. The spin part of this system is defined by $15$
parameters: $6$ of them represent Bloch vectors for the first and the second
qubits respectively. The other nine parameters represent the component of
the correlation tensor \cite{metwally}.

\begin{equation}  \label{qubit1}
\rho_{ab}=\frac{1}{4}\left(1+\sum_{i=1}^3{s_i\sigma_i}+\sum_{j=1}^{3}{%
t_j\tau_j}+ \sum_{ij=1}^{3}{c_{ij}\sigma_i\tau_j}\right),
\end{equation}
where $\sigma_i,\tau_j$, $i,j=1,2,3$ are the Pauli matrices for the first
and second's qubit respectively, $s_i, t_j$ with $s_i=tr\{\rho_{ab}
\sigma_i\}$ and $t_j=tr\{\rho_{ab} \tau_j\}$ are Bloch vectors for both
qubits respectively. The tensor correlation is defined by a $3\times 3$
matrix with elements are defined by $c_{ij}=tr\{\rho_{ab}\sigma_i\tau_j\}$.
For example, $c_{11}=tr\{\rho_{ab}\sigma_1\tau_1\},c_{x12}=tr\{\rho_{ab}%
\sigma_1\tau_2\}, c_{13}=tr\{\rho_{ab}\sigma_1\tau_3\}$ and so on. From the
general form (\ref{qubit1}), one can obtains different classes \cite%
{metwally1}.

The dynamics of the state (\ref{qubit1}) under the effect of the lorentz
transformation 
is characterized by its new Bloch vectors,
\begin{eqnarray}
\tilde{s}_{1} &=&s_{1}\cos \phi -s_{2}\sin \phi ,\quad \tilde{s}%
_{2}=s_{2}\cos \phi -s_{1}\sin \phi ,\quad \tilde{s}_{3}=\frac{s_{3}}{2}%
(1+\cos 2\phi )+\frac{t_{3}}{2}(1-\cos 2\phi ),  \nonumber \\
\tilde{t}_{1} &=&t_{1}\cos \phi -t_{2}\sin \phi ,\quad \tilde{t}%
_{2}=t_{1}\sin \phi +t_{2}\cos \phi ,\quad \tilde{t}_{3}=\frac{s_{3}}{2}%
(1-\cos 2\phi )+\frac{t_{3}}{2}(1+\cos 2\phi ),  \nonumber \\
&&
\end{eqnarray}%
and the $9$ elements of the cross dyadic which are defined as,
\begin{eqnarray}
\tilde{c}_{xx} &=&\frac{1}{2}(1+\cos 2\phi )c_{xx}+\frac{1}{2}(1-\cos 2\phi
)c_{yy}+\frac{1}{2}(c_{xy}-c_{yx})\sin 2\phi ,  \nonumber \\
\tilde{c}_{xy} &=&\frac{1}{2}(1+\cos 2\phi )c_{xy}+\frac{1}{2}(1-\cos 2\phi
)c_{yx}-\frac{1}{2}(c_{xx}+c_{yy})\sin 2\phi ,  \nonumber \\
\tilde{c}_{xz} &=&\cos \phi ~c_{xz}-\sin \phi ~c_{yz},  \nonumber \\
\tilde{c}_{yx} &=&\frac{1}{2}(1+\cos 2\phi )c_{yx}+\frac{1}{2}(1-\cos 2\phi
)c_{xy}+\frac{1}{2}(c_{xx}+c_{yy})\sin 2\phi ,  \nonumber \\
\tilde{c}_{yy} &=&\frac{1}{2}(1+\cos 2\phi )c_{yy}-\frac{1}{2}(1-\cos 2\phi
)c_{xx}-\frac{1}{2}(c_{xy}+c_{yx})\sin 2\phi ,  \nonumber \\
\tilde{c}_{yz} &=&\cos \phi ~c_{yz}+\sin \phi ~c_{xz},  \nonumber \\
\tilde{c}_{zx} &=&\cos \phi ~c_{zx}+\sin \phi ~c_{zy},\quad \quad \tilde{c}%
_{zy}=\cos \phi ~c_{zy}+\sin \phi ~c_{zx},~\quad \tilde{c}_{zz}=c_{zz}.
\end{eqnarray}%
In the following section we shall consider some important example in context
of quantum information as Werner state, $x-$ state and a generic pure state,
were we investigate the dynamics of these states under the effect of Lorentz
transformation. Also, we quantify the degree of entanglement of the new
states.

\section{Entanglement}

In this section, we quantify the degree of entanglement under the effect of
Lorentz transformation. We use Wootters'concurrence \cite{Woottors}, for
this propose, which is defined as,
\begin{equation}
\mathcal{E}=\max\Bigl\{0,\lambda_1-\lambda_2-\lambda_3-\lambda_4\Bigr\},
\end{equation}
where $\lambda_1\geq\lambda_2\geq\lambda_3\geq\lambda_4$ and $\lambda_i$ are
the eigenvalues of the density operator $\rho=\sigma_2\tau_2\rho^*\sigma_2%
\tau_2$, $\rho^*$ is the complex conjugate of $\rho$.

\begin{figure}[tbp]
\begin{center}
\includegraphics[width=15pc,height=12pc]{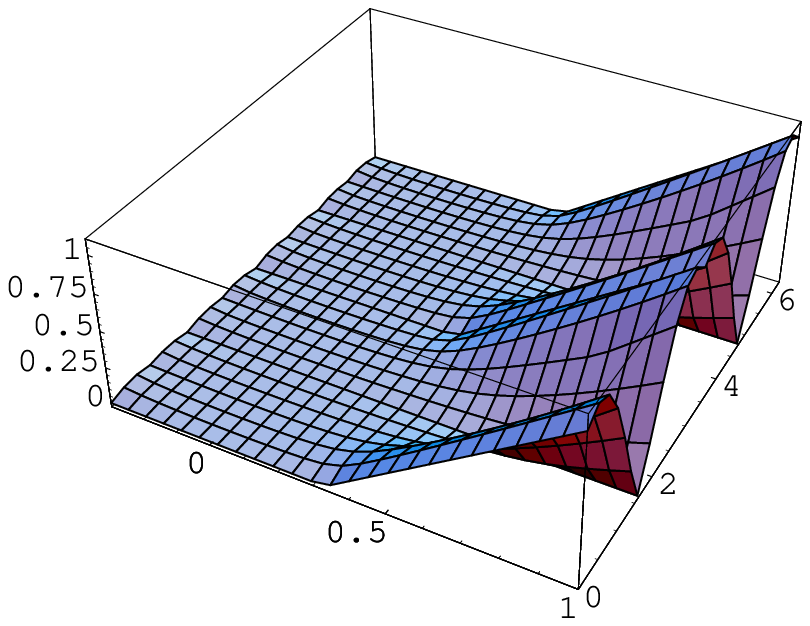} \put(40,20){$x$}
\put(-150,20){$x$} \put(-190,70){$\mathcal{E}$} \put(-6,70){$\mathcal{E}$}
\put(-15,40){$\mathcal{\varphi}$} \put(170,40){$\mathcal{\varphi}$} %
\includegraphics[width=15pc,height=12pc]{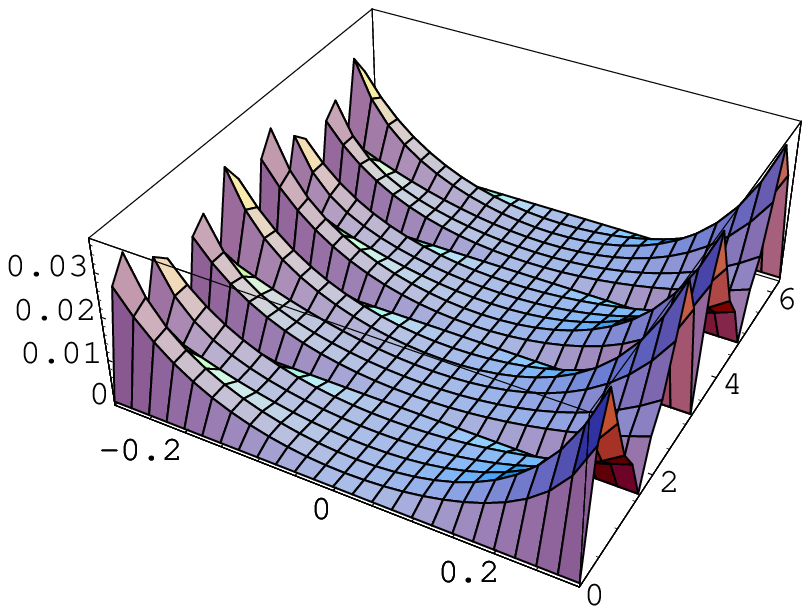} \put(-100,10){$x$}
\put(-190,70){$\mathcal{E}$} \put(-10,40){$\mathcal{\varphi}$}
\end{center}
\caption{The dynamics of entanglement under Lorentz transformation for a
system is initially prepared in different classes of Werner states}
\end{figure}

\begin{enumerate}
\item {X-state}: This class of states \cite{Eberly} represents one of most
important classes in the context of quantum information \cite%
{Chen,Ali,Rau,Alaa}. On the other hand, from this class, one can generate
Werner state and Bell states. The density operator of this class is given
by,
\begin{equation}  \label{x-state}
\rho_x=\frac{1}{4}(1+c_{xx}\sigma_x\tau_x+c_{yy}\sigma_y\tau_y+c_{zz}%
\sigma_z\tau_z)
\end{equation}
Under Lorentz transformation, this state (\ref{x-state}) is transformed into
\begin{eqnarray}
\tilde\rho_x&=&\frac{1}{4}\Bigl\{1+\frac{1}{2}\{(1+\cos2\phi)c_{xx}+(1-\cos2%
\phi)c_{yy}\}\sigma_x\tau_x -\frac{1}{2}(c_{xx}+c_{yy})sin2\phi(
\sigma_x\tau_y-\sigma_y\tau_x)  \nonumber \\
&& \hspace{1cm} -\frac{1}{2}\{(1-\cos2\phi)c_{xx}-(1+\cos2\phi)c_{yy}\}%
\sigma_y\tau_y+c_{zz}\sigma_z\tau_z\Bigl\}
\end{eqnarray}
Fig.(1) shows the behavior of entanglement of Werner state \cite{Wer}, where
this state is initially defined by its zero Bloch vectors, i.e., $%
\mathord{\buildrel{\lower3pt\hbox{$\scriptscriptstyle\rightarrow$}}\over s}=%
\mathord{\buildrel{\lower3pt\hbox{$\scriptscriptstyle\rightarrow$}}\over t}%
=0 $ and $c_{xx}=c_{yy}=c_{zz}=-x$. It has been shown that has been shown
that this state is separable for $x\in[-\frac{1} {3} ,\frac{1}{3} ]$ and
nonseparable for $\frac{1} {3} < x \leq 1$ (see for example \cite{metally}).

For entangled state classes i.e., $x\in[\frac{1}{3},1]$, the entanglement
decreases to reach its minimum value as $\varphi$ increases. Then the
entanglement re-birthed again to reach its maximum bound. This maximum bound
depends on the value of the parameter $x$, where Werener state represents a
singlet state (Bell state) at $x=1$, which is a maximum entangled state.

On the other hand, for separable classes namely for any $x\in[-\frac{1} {3} ,%
\frac{1}{3} ]$, the separable states turns into entangled states as shown in
Fig. 1b. However the degree of entanglement is smaller than those depicted
for initially entangled classes.

\begin{figure}[tbp]
\begin{center}
\includegraphics[width=22pc,height=14pc]{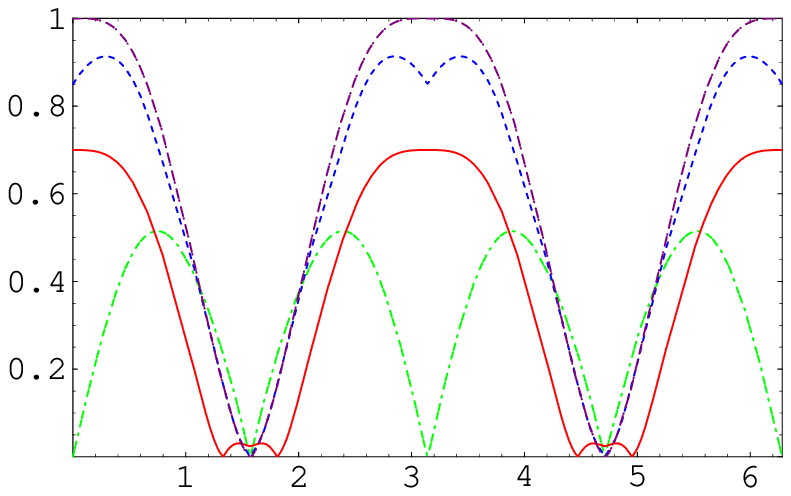} \put(-130,-8){$\varphi$}
\put(-270,80){$\mathcal{E}$}
\end{center}
\caption{The dynamics of entanglement under Lorentz transformation for a
system is initially prepared in maximum entangled states (dash-dot curve),
Werner state with $x=-0.6$ (solid curves) and $X$-states with (dot ) curves $%
c_{xx}=-0.9, c_{yy}=-0.8$ and $c_{zz}=-0.7$.}
\end{figure}
Fig.(2) shows the effect of Lorentz transformation on the degree of
entanglement for maximum entangled state, where $c_{xx}=c_{yy}=c_{zz}=-1$,
x-state which defined by $c_{xx}=-0.9\neq c_{yy}=-0.8\neq c_{zz}=-0.7$ and
Werner state with $c_{xx}=c_{yy}=c_{zz}=0.7$. As a general behavior, the
entanglement $\mathcal{E}$ decreases as $\phi$ increase. The decreasing rate
depends on the degree of entanglement for the initial state. However for
maximum entangled sates, MES the entanglement decreases very fast to
completely vanish and then re-birth again to reach its maximum value $(%
\mathcal{E}=1$). For less entangled states, a similar behavior is depicted
as MES, but the entanglement is complectly death for longer interval of $%
\phi $. Also, this figure shows the effect of Lorentz transformation on
systems prepared initially in a separable states, where we set $x=-0.6$. It
is clear that the initial entanglement is zero. However as $\varphi$
increases an entangled state is generated and its maximum value is $\mathcal{%
E}\simeq 0.44 $.

\item {Generic pure state}

This state is described by the density operator\cite{metwally},
\begin{equation}
\rho _{p}=\frac{1}{4}(1+p(\sigma _{x}-\tau _{x})-\sigma _{x}\tau
_{x}-q(\sigma _{y}\tau _{y}+\sigma _{z}\tau _{z})),  \label{Final-p}
\end{equation}%
where, $q=\sqrt{1-p^{2}}$ . Under the Lorentez transformation this state is
transformed into
\begin{eqnarray}
\tilde{\rho}_{p} &=&\frac{1}{4}\Bigl\{1+p\cos \phi (\sigma _{x}-\tau _{x})-%
\frac{1}{2}\{(1+q)+(1-q)\cos 2\phi \}\sigma _{x}\tau _{x}+\frac{1}{2}%
(1+q)\sin 2\phi \sigma _{x}\tau _{y}  \nonumber \\
&-&\frac{1}{2}(1-q)\sin 2\phi \sigma _{y}\tau _{x}+\frac{1}{2}%
\{(1-q)-(1+q)\cos 2\phi \}\sigma _{y}\tau _{y}-q\sigma _{z}\tau _{z}\Bigr\}.
\end{eqnarray}

\begin{figure}[tbp]
\begin{center}
\includegraphics[width=20pc,height=15pc]{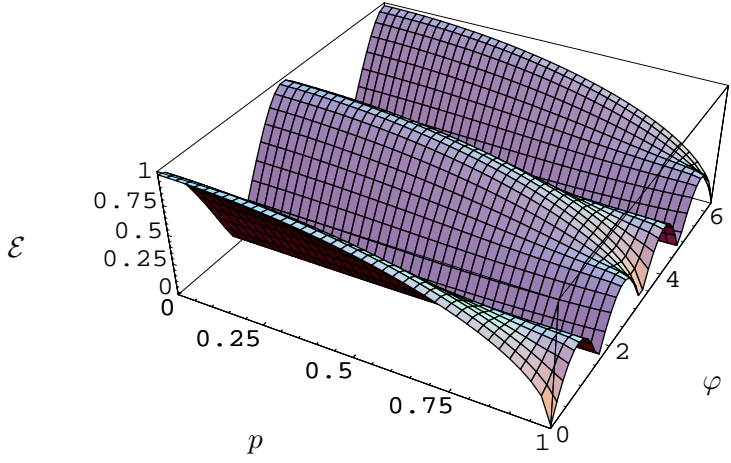} \put(-180,8){$p$}
\put(-270,80){$\mathcal{E}$} \put(-10,30){$\varphi$}
\end{center}
\caption{The same as Fig.(1) but for a system is prepared initially in a
pure state.}
\end{figure}

\begin{figure}[tbp]
\begin{center}
\includegraphics[width=22pc,height=14pc]{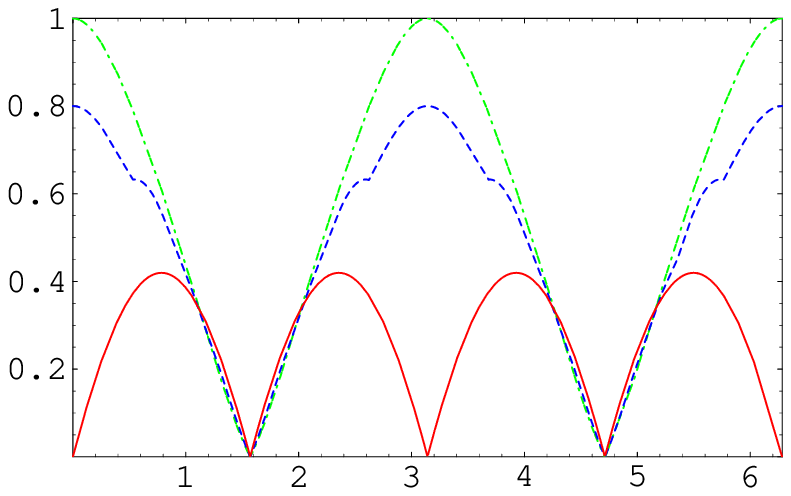} \put(-130,-8){$\varphi$}
\put(-270,80){$\mathcal{E}$}
\end{center}
\caption{The dynamics of Entanglement $\mathcal{E}$ for a system prepared
initially in a pure state. The dash-dot curve for $p=0$ (MES), dot curve for
PES with $p=0.6$ and the solid curve for separable state i.e., $p=1$. }
\end{figure}
\end{enumerate}

\section{Information loss}

In this section, we investigate the effect of Lornetz transformation on the
local and non local information via calculating the entropy of both
subsystems and the total state. The entropy of a a density operator $\rho$
is defined by Von Numman entropy $\mathcal{P}_n=-\varrho ln\varrho$. This
value indicate how much of information that the state $\varrho$ is lost.

In Fig. 4, we investigate the dynamics of entropy of different classes of
Werner type under the effect of Lorentz transformation. It is clear that
starting from entangled classes i.e., $\frac{1}{3}<x\leq 1$, the initial
entropy is not maximum. This means that these state contains some quantum
information. As $x$ increases the entropy $\mathcal{P}_{n}$ decreases to
reach its minimum values at $x=1$ i.e., the initial state is maximum
entangled state. On the other hand, as one increases $\varphi $, the entropy
increases faster for larger values of $x$, namely classes with larger degree
of entanglement. However for less entangled states the entropy is slightly
increases. This show that the less entangled classes are more robust under
Lorentz transformation. These results are shown in Fig. 4a. Starting from a
separable classes, the initial entropy $\mathcal{P}_{n}$ is larger than that
for entangled states. However, the entropy reaches its maximum value for
less entangled states as shown in Fig. $4b$. Also, as one increases $\varphi
$, the entropy $\mathcal{P}_{n}$ oscillates between its maximum and minimum
values.

\begin{figure}[tbp]
\begin{center}
\includegraphics[width=15pc,height=12pc]{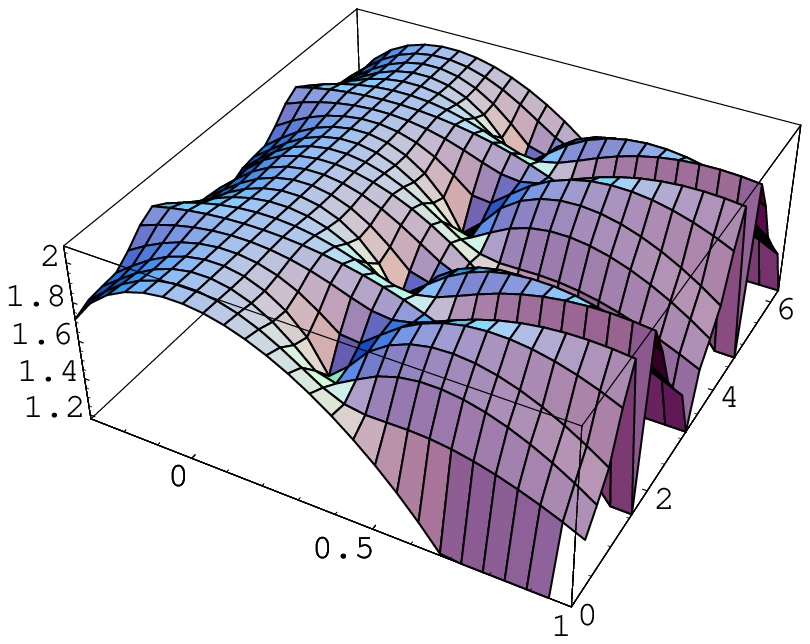}~\quad\quad \put(-150,20){%
$x$} \put(20,20){$x$} \put(-190,70){$\mathcal{P}_n$} \put(-6,70){$\mathcal{P}%
_n$} \put(-15,40){$\mathcal{\varphi}$} \put(170,40){$\mathcal{\varphi}$} %
\includegraphics[width=15pc,height=12pc]{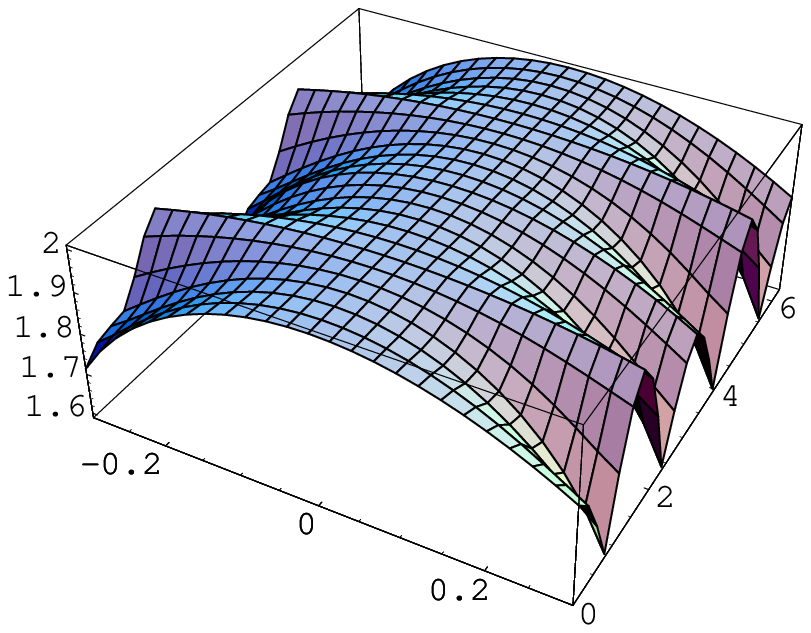}
\put(-100,10){$x$} \put(-190,70){$\mathcal{P}_n$} \put(-10,40){$\mathcal{%
\varphi}$}
\end{center}
\caption{The behavior of entropy $\mathcal{P}_n$ under Lorentz
transformation for different classes of Werner state}
\end{figure}

\begin{figure}[tbp]
\begin{center}
\includegraphics[width=22pc,height=14pc]{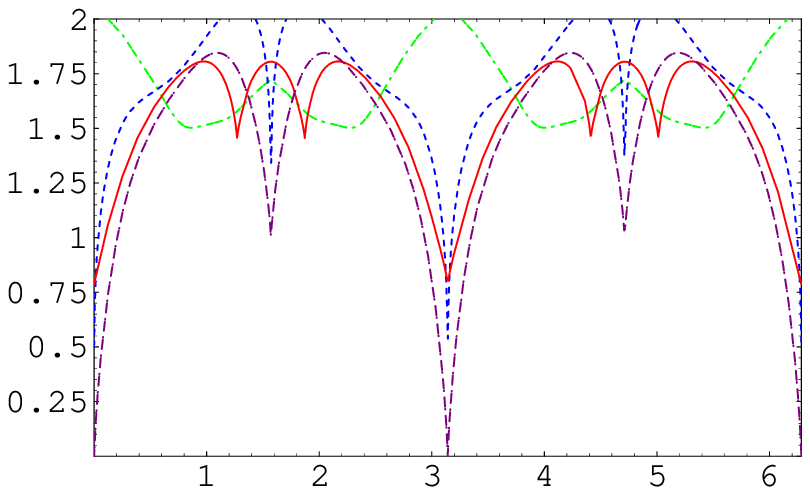} \put(-130,-10){$x$}
\put(-270,80){$\mathcal{P}_n$}
\end{center}
\caption{The behavior of Entropy $\mathcal{P}_n$ under Lorentz
transformation for a system is initially prepared in maximum entangled
states (dash- curve), Werner state with $x=-0.6$ (dash-dot) and $x=0.7$ (dot
curve) and $X$-states with (solid ) curves $c_{xx}=-0.9, c_{yy}=-0.8$ and $%
c_{zz}=-0.7$ (solid-curve).}
\end{figure}

Fig. 5 shows the behavior of the entropy of different classes of the $x-$
state (\ref{x-state}). It is clear that, starting from a maximum entangled
class, i.e., we set $c_{xx}=c_{yy}=c_{zz}=-1$, the entropy $\mathcal{P}_n=0$
at $\varphi=0$. This means that the amount of information on this state is
maximum. However as one increases $\varphi$, the entropy increases on the
expanse of the non-local information to reach its maximum values $(\mathcal{P%
}_n=2)$. This maximum value is reached at $\varphi\simeq\frac{\pi}{3}$ i.e.,
at the same value of the minimum amount of entanglement (see Fig. 2). Also,
this figure depicts the behavior of entropy for a class of Werner state,
where we set $c_{xx}=c_{yy}=c_{zz}=x=0.7$ i.e., the initial state represents
a partial entangled state with small degree of entanglement, the initial
value of entropy is larger than that depicted for maximum entangled state.
As one increases the Lornetz transformation's parameter $\varphi$, the
entropy decreases to reach its minimum value and increases again. However
for this class the minimum value always larger than the initial values. The
behavior of entropy starting from a separable class of Werner type, where we
set $x=-0.6$, the initial entropy is maximum, i.e., $\mathcal{P}_n=2$.
However this entropy oscillates between minimum and maximum values as $%
\varphi$ increases. This show that there is an entangled state is generated
for different values of $\varphi$. Finally this figure depict the behavior
of entropy under the effect of Lorentz transformation for a class of $x-$
states i.e., $c_{xx}\neq c_{yy}\neq c_{zz}$. It is clear that a similar
behavior is shown as the pervious classes, but the entropy doesn't reach its
maximum value. So, this class is more robust than the previous class under
the effect of Lorentz transformation.

\section{Conclusion}

In this contribution obtain an analytical form for the spin part of the a
general two-qubit systems. Some classes as Werner, $x-$ and a generic pure
states are investigated extensively. The behavior of entanglement as well as
the entropy which measures the information loses are investigated. It is
shown that, the degree of entanglement decreases faster to completely vanish
for system s prepared initially in maximum entangled states. Starting from
less entangled states, the entanglement decreases gradually to reach its
minimum value. On the other hand, the entanglement rebirth faster for the
systems prepared initially in maximum entangled states. As one increases the
lorentz parameter, the re-birth entanglement doesn't exceeds its initial
values. Our results show that one can generate entangled states starting
from systems prepared initially in a separable states. In this case the
generated entangled states depends on the structures of the initial systems.
It is clear that, starting from a separable state generated from a pure
state has a larger degree of entanglement compared with that obtained from
Werner classes. The information loss is quantified by investigating the
behavior of entropy for different classes of initial states. It is clear
that the entropy of systems prepared initially in a partially entangled
states increases gradually, but it increases fast for systems prepared
initially in maximum entangled states. Therefore partially entangled states
are more robust than maximum entangled states and consequently the rate of
information loss is larger for maximum entangled states compared with that
for partially entangled states.

\end{document}